\documentclass[twocolumn,showpacs,preprintnumbers,amsmath,amssymb,superscriptaddress]{revtex4}

\usepackage{amsfonts,amsmath,amssymb,graphicx}
\usepackage{amsfonts}
\usepackage{amsmath}
\usepackage{amssymb}
\usepackage{graphicx}

\def\kbar{\protect\@kbar}
\def\@kbar{\relax \bgroup
\def\@tempa{\hbox{\raise.73\ht0
\hbox to0pt{\kern.25\wd0\vrule width.5\wd0 height.1pt
depth.1pt\hss}\box0}}\mathchoice{\setbox0\hbox{$\displaystyle
k$}\@tempa}{\setbox0\hbox{$\textstyle
k$}\@tempa}{\setbox0\hbox{$\scriptstyle
k$}\@tempa}{\setbox0\hbox{$\scriptscriptstyle k$}\@tempa}\egroup}

\begin{document}

\title{\textbf{Experimental Realization of Quantum-Resonance Ratchets at
Arbitrary Quasimomenta}}
\author{I. Dana}
\affiliation{Minerva Center and Department of Physics, Bar-Ilan
University, Ramat-Gan 52900, Israel}
\author{V. Ramareddy}
\affiliation{Department of Physics, Oklahoma State University,
Stillwater, Oklahoma 74078-3072, USA}
\author{I. Talukdar}
\affiliation{Department of Physics, Oklahoma State University,
Stillwater, Oklahoma 74078-3072, USA}
\author{G. S. Summy}
\affiliation{Department of Physics, Oklahoma State University,
Stillwater, Oklahoma 74078-3072, USA}

\begin{abstract}
\noindent Quantum-resonance ratchets associated with the kicked
particle are experimentally realized for \emph{arbitrary
quasimomentum} using a Bose-Einstein condensate (BEC) exposed to a
pulsed standing light wave. The ratchet effect for general
quasimomentum arises even though both the standing-wave potential
and the initial state of the BEC have a point symmetry. The
experimental results agree well with theoretical ones which take
into account the finite quasimomentum width of the BEC. In
particular, this width is shown to cause a \emph{suppression} of the
ratchet acceleration for exactly \textquotedblleft
resonant\textquotedblright\ quasimomentum, leading to a saturation
of the directed current. The finite-width ratchet effect is
generally similar to the ideal one in several aspects.
\end{abstract}

 \maketitle

Understanding quantum transport in classically chaotic systems is a
problem of both fundamental and practical importance. A wide variety
of interesting quantum-transport phenomena have been discovered in
simple but representative quantized models of Hamiltonian dynamics
\cite{qc}. These phenomena either exhibit fingerprints of classical
chaotic transport in a semiclassical regime or are purely quantum in
nature. A paradigmatic and realistic class of model systems which
has been studied most extensively, both theoretically and
experimentally, consists of the periodically kicked rotor and
variants of it \cite
{qc,qr,ao,qam,qam1,kp,kp1,dd,hqr,crd,qrae,qrr,dr,eqr0}. These
systems feature some of the most well-known phenomena in the field
of quantum chaos, such as dynamical localization \cite{qc}, i.e.,
the quantum suppression of classical chaotic diffusion, and the
diametrically opposite phenomenon of \emph{quantum resonance} (QR)
\cite{qr}. The latter is a purely quantum quadratic growth of the
mean kinetic energy in time occurring for special values of an
effective Planck constant. The experimental realization of the
kicked rotor using atom-optics techniques \cite{ao} has led to
breakthroughs in the study of quantum chaos. Such experiments
actually realize the kicked particle, not the kicked rotor, since
atoms move on lines and not on circles like a rotor. However, the
two systems can be exactly related due to the conservation of the
particle \emph{quasimomentum} $\beta $ \cite{kp,dd} (see also
below). General manifestations of QR in the kicked particle for
arbitrary $\beta $ \cite{kp,kp1,dd} have been observed in
atom-optics experiments \cite{kp1,hqr}.\newline

An important concept introduced recently in classical and quantum
Hamiltonian transport is that of a \textquotedblleft
ratchet\textquotedblright . This is a spatially periodic system in
which, without a biased force, a directed current of particles can
be established. Ratchet models were originally proposed as
mechanisms for some kinds of biological motors and as nanoscale
devices for several applications \cite{ra}. In these and other
\cite{ra1} contexts, the directed current is due to a
spatial/temporal asymmetry combined with noise and dissipation. In a
classical Hamiltonian system, dissipation is absent and noise is
replaced by deterministic chaos. Here a directed current of
particles in the chaotic sea may arise under asymmetry conditions
for a mixed phase space \cite{cra}. The corresponding quantized
system may exhibit a significant ratchet behavior even in a fully
chaotic regime \cite{crd,qrae,qra1}. Such a behavior, which occurs
in a variant of the kicked rotor and can be related to the
underlying classical dynamics, was observed recently in experiments
using ultra-cold atoms \cite{qrae}. Under \emph{exact} QR
conditions, theory predicts a purely quantum ratchet
\emph{acceleration}, i.e., a linear increase of the directed current
(the mean momentum) in time \cite{qrr,dr}. An experimental
observation of this phenomenon on a short time interval was reported
quite recently for the usual kicked rotor \cite{eqr0}, with $\beta
=0$. It is known, however, that the unavoidable experimental
uncertainty in $ \beta $\ strongly affects QR \cite{kp,kp1,dd} and
it is therefore natural to ask how it will affect the QR-ratchet
acceleration.\newline

In this Letter, we present an experimental realization of quantum
ratchets associated with QR of the kicked particle for
\emph{arbitrary} values of the quasimomentum $\beta $. The
experiments are conducted by exposing a Bose-Einstein condensate
(BEC) to a pulsed optical standing wave which creates a potential
$V(x)=V_{\mathrm{\max }}\cos (Gx-\gamma )$. Here $x$ is the spatial
coordinate along the standing wave, $G=4\pi /\lambda $ is the
\textquotedblleft grating vector\textquotedblright , $\lambda $ is
the light wavelength, and $\gamma $ is an \textquotedblleft
offset\textquotedblright\ phase. The BEC is prepared in an initial
state $\psi _{0}(x)$ with $ \left\vert \psi _{0}(x)\right\vert
\propto \left\vert \cos (Gx/2)\right\vert $, very different from the
pure momentum state normally used in quantum-chaos experiments
\cite{ao,qam,qam1,kp1,hqr,qrae}. In fact, $\psi _{0}(x)$ is the
simplest state for which QR-ratchet effects can be observed
\cite{dr}. Our experimental results agree well with a very recent
general theory of QR ratchets \cite{dr}, after including in this
theory the effect of the finite quasimomentum width $\Delta \beta $
of the BEC. This width is shown to cause a \emph{suppression} of the
QR-ratchet acceleration for exactly \textquotedblleft
resonant\textquotedblright\ $\beta $ \cite{dr}, leading to a
saturation of the directed current. For general $\beta$ and
$\gamma$, the finite-width ratchet effect is similar to the ideal
(zero-width) one in several aspects. Unlike other kinds of quantum
ratchets \cite{crd,qrae,cra,qra1}, the QR ratchet exhibits fully
\emph{symmetric} features; both $V(x)$ and $\psi _{0} (x)$ have a
point symmetry around some center. The directed current for general
$\beta $ depends strongly on the relative displacement between the
two symmetry centers, here given by $\gamma $. Figure 1 shows a plot
of the resonant current after 5 standing-wave pulses as a function
of $\gamma $. Note that the current direction can be easily
controlled by $\gamma $ and that the current vanishes for $\gamma
=0$ (coinciding symmetry centers).\newline

\begin{figure}
  \includegraphics[width=6cm]{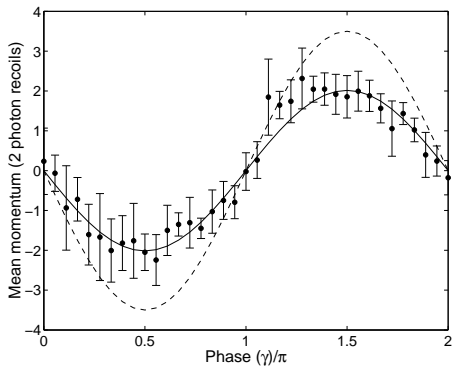}\\
  \caption{Mean momentum vs. phase angle $\protect\gamma $ for kicking
strength $k=1.4$, kick number $t=5$, and resonant quasimomentum
$\protect \beta =0.5$. The filled circles and associated error bars
are from the experiment. The solid line corresponds to the theory
(\protect\ref{ARe}) for a BEC with width $\Delta \protect\beta
=0.056$. The dashed line corresponds to the ideal ($\Delta
\protect\beta =0$) theory (\protect\ref{Nae}).}\label{fig1}
\end{figure}

We start by summarizing the basic theory (see more details in
Refs.\thinspace \cite{kp,dd,dr}). Using dimensionless quantities,
the quantum kicked particle is described by the general Hamiltonian
$\hat{H}= \hat{p}^{2}/2+kV(\hat{x})\sum_{t}\delta (t^{\prime }-t\tau
)$, where $(\hat{p },\hat{x})$ are momentum and position operators,
$k$ is the kicking strength, $V(x)$ is an arbitrary periodic
potential, $\tau $ is the kicking period, and $t^{\prime }$ and $t$
are the continuous and \textquotedblleft integer\textquotedblright\
times. The units are chosen so that the particle mass is $1$, $\hbar
=1$, and the period of $V(x)$ is $2\pi $. The translational
invariance of $\hat{H}$ in $\hat{x}$ implies the conservation of the
quasimomentum $\beta $, $0\leq \beta <1$, in the time evolution of a
Bloch wave packet $\exp (i\beta x)\psi (x)$. Here $\psi (x)$ is a
$2\pi$-periodic function, so that one can consider $x$ as an angle
$\theta $. Introducing the corresponding angular-momentum operator
$\hat{N}=-id/d\theta $, the time evolution of $\psi (\theta )$ in
one period $\tau $ is given by $ \hat{U}_{\beta }\psi (\theta )$,
where
\begin{equation}
\hat{U}_{\beta }=\exp \left[ -ikV(\hat{\theta})\right] \exp \left[ -i(\tau
/2)\left( \hat{N}+\beta \right) ^{2}\right] .  \label{Ub}
\end{equation}
The evolution operator (\ref{Ub}) describes a \textquotedblleft
$\beta$-kicked-rotor\textquotedblright\ which is related to the
kicked particle at fixed $\beta $ by $\hat{p}=$ $\hat{N}+\beta $,
i.e., $\beta $ is the \textquotedblleft
fractional\textquotedblright\ part of the particle momentum. QR of a
$\beta $-kicked-rotor is a quadratic growth of the mean kinetic
energy at sufficiently large times $t$: $\langle \psi
_{t}|\hat{N}^{2} /2|\psi _{t}\rangle \sim Dt^{2} $, where $\psi
_{t}(\theta )=\hat{U}_{\beta }^{t}\psi _{0}(\theta )$ is an evolving
wave packet and $D$ is some coefficient. This behavior will
generally occur provided two conditions are satisfied \cite{dd}: (a)
The effective Planck constant $\tau /(2\pi )$ is a rational number
$l_{0}/q_{0}$, where $(l_{0},q_{0})$ are coprime integers; (b)
$\beta $ is also a rational number $\beta _{\mathrm{r}}$
characterized by two integers $(r,g)$ defined as follows: $g$ is the
smallest integer such that $r=(\beta _{\mathrm{r}}+gq_{0}/2)gl_{0}$
is integer. Then, the resonant quasimomentum $\beta =\beta
_{\mathrm{r}}$ can be written as $\beta =\beta _{r,g}$.\newline

We shall focus on the case of integer $\tau /(2\pi )=l_{0}$
($q_{0}=1$), corresponding to the \textquotedblleft
main\textquotedblright\ QRs. In this case, the time evolution of
wave packets $\psi _{t}(\theta )$ under (\ref{Ub}) can be exactly
calculated for arbitrary potential $V(\theta )=\sum_{m}V_{m}\exp
(-im\theta )$ and for any $\beta $ \cite{dd}. One can then evaluate
the current $\langle \hat{p}\rangle _{t}$, i.e., the mean momentum
of the evolving Bloch wave. Using $\langle \hat{p}\rangle
_{t}=\langle \hat{N} \rangle _{t}+\beta $, where $\langle
\hat{N}\rangle _{t}\equiv \langle \psi _{t}|\hat{N}|\psi _{t}\rangle
$, the \emph{change} in the current relative to its initial value is
given by $\Delta \langle \hat{p}\rangle _{t}=\langle \hat{N}\rangle
_{t}-\langle \hat{N}\rangle _{0}$. This change is a measure of the
ratchet effect induced by the kicking. Exact results for $\Delta
\langle \hat{p}\rangle _{t}$ can be derived for general $\psi
_{0}(\theta )$ \cite{dr}, but we shall restrict ourselves here to
the simplest initial wave packet giving ratchet effects: $\psi
_{0}(\theta )=\left[ 1+\exp (-i\theta ) \right] /\sqrt{4\pi }$. The
arbitrary potential $V(\theta )$ can be written, up to a real
factor, as $V(\theta )=\cos (\theta -\gamma )+V_{2}(\theta )$, where
$V_{2}(\theta )$ contains only harmonics of order $m\geq 2$. Then,
by applying the general exact result (17) in Ref. \cite{dr} to our
system (corresponding to $w=0$ and $T=1$ in \cite{dr}), we find for
the initial wave packet above that
\begin{equation}
\Delta \langle \hat{p}\rangle _{t}=\frac{k}{2}\frac{\sin (\tau _{\beta }t/2)
}{\sin (\tau _{\beta }/2)}\sin \left[ (t+1)\tau _{\beta }/2-\gamma \right] ,
\label{Nae}
\end{equation}
where $\tau _{\beta }=\pi l_{0}(2\beta +1)$. Due to the simple
choice of $ \psi _{0}(\theta )$, this result is completely
\emph{independent} of $ V_{2}(\theta )$. Thus, without loss of
generality, henceforth we take $ V(\theta )=\cos (\theta -\gamma )$.
The denominator in (\ref{Nae}) vanishes if $\tau _{\beta }=2\pi r$
for some integer $r$. This corresponds precisely to a resonant value
of $\beta =\beta _{r,g}=r/l_{0}-1/2\ \ \mathrm{mod}(1)$, with $g=1$.
For $\beta =\beta _{r,1}$, Eq. (\ref{Nae}) reduces to a linear
growth in time $t$, a ratchet acceleration: $\Delta \langle
\hat{p}\rangle _{t,r}=-(k/2)\sin (\gamma )$\thinspace $t$.\newline

In an experimental realization of QR ratchets using a kicked BEC,
the small but finite initial momentum width of the BEC can be
important. Here we consider a mixture of quasimomenta $\beta
^{\prime }$, having a Gaussian distribution with average $\beta $
and standard deviation $\Delta \beta $: $\Gamma _{\beta ,\Delta
\beta }(\beta ^{\prime })=(\Delta \beta \sqrt{2\pi } )^{-1}\exp
\left\{ -(\beta ^{\prime }-\beta )^{2}/\left[ 2(\Delta \beta )^{2}
\right] \right\} $. For small $\Delta \beta $, this is a good
approximation of the actual initial momentum distribution, resulting
from the gradient of the mean-field energy of the condensate
\cite{Ketterle}. The average of (\ref{Nae}) over $\beta =\beta
^{\prime }$ with distribution $\Gamma _{\beta ,\Delta \beta }(\beta
^{\prime })$ can be exactly calculated:
\begin{equation}
\left\langle \Delta \langle \hat{p}\rangle _{t}\right\rangle _{\Delta \beta
}=\frac{k}{2}\sum_{s=1}^{t}\sin \left( \tau _{\beta }s-\gamma \right) \exp
\left[ -2\left( \pi l_{0}\Delta \beta s\right) ^{2}\right] .  \label{ANae}
\end{equation}
Unlike (\ref{Nae}), the expression (\ref{ANae}) tends to a
well-defined finite value as $t\rightarrow \infty $, for \emph{all}
$\beta $. In particular, for resonant $\beta =\beta _{r,1}$ with
$\tau _{\beta }=2\pi r$, Eq. (\ref{ANae}) reduces to
\begin{equation}
\left\langle \Delta \langle \hat{p}\rangle _{t}\right\rangle _{r,\Delta
\beta }=-\frac{k}{2}\sin (\gamma )\sum_{s=1}^{t}\exp \left[ -2\left( \pi
l_{0}\Delta \beta s\right) ^{2}\right] .  \label{ARe}
\end{equation}
The result (\ref{ARe}) implies a suppression of the ratchet
acceleration above, which is recovered as $\Delta \beta \rightarrow
0$. In practice, for sufficiently small $\Delta \beta $, the value
of $\left\langle \Delta \langle \hat{p}\rangle _{t}\right\rangle
_{\Delta \beta }$ for $\beta $ close to $\beta _{r,1}$ is much
larger than that for generic $\beta $, except when $|\sin (\gamma
)|$ is very small.\newline

It should be noted that both the potential $V(\theta )=\cos (\theta
-\gamma ) $ and the initial wave packet $\psi _{0}(\theta )=\left[
1+\exp (-i\theta ) \right] /\sqrt{4\pi }$ have a point symmetry:
$V(2\gamma -\theta )=V(\theta ) $ (inversion around $\theta =\gamma
$) and $\psi _{0}^{\ast }(-\theta )=\psi _{0}(\theta )$ (inversion
around $\theta =0$ accompanied by time reversal). The results
(\ref{Nae})-(\ref{ARe}) depend on the relative displacement $\gamma
$ between the symmetry centers of $V(\theta )$ and $ \psi
_{0}(\theta ) $. For $\gamma =0$, the system is \textquotedblleft
symmetric\textquotedblright\ and there is no resonant ratchet
effect, $ \left\langle \Delta \langle \hat{p}\rangle
_{t}\right\rangle _{r,\Delta \beta }=0$. At fixed $k$,
$|\left\langle \Delta \langle \hat{p}\rangle _{t}\right\rangle
_{r,\Delta \beta }|$ is largest for $\gamma =\pm \pi /2$, values of
$\gamma $\ which may be viewed as corresponding to \textquotedblleft
maximal asymmetry\textquotedblright\ situations. The direction of
the change (\ref{ARe}) in the current is given by the sign of $
-\sin (\gamma )$. We shall henceforth use $l_{0}=1$, corresponding
to the \textquotedblleft half-Talbot time\textquotedblright , so
that the only resonant value of $\beta $ is $\beta =0.5$. It is easy
to see that $\langle \hat{p}\rangle _{0}=\langle \hat{N}\rangle
_{0}+\beta =0$ for $\beta =0.5$ and then $\Delta \langle
\hat{p}\rangle _{t}=\langle \hat{p}\rangle _{t}$.
\newline

Our experiments were carried out using the all-optical BEC apparatus
described in Ref. \cite{qam1}. After creating a BEC of $\sim 50000$
$^{\text{ 87}}$Rb atoms in a focused CO$_{2}$ laser beam, we applied
a series of optical standing-wave pulses from a diode laser beam
(6.8 GHz red detuned from the 780 nm laser cooling transition)
propagating at $52^{\text{\textrm{o }}}$ to the vertical. Through an
ac-Stark shift, the standing wave changed the energy of the atoms by
an amount proportional to the light intensity. The resulting
spatially periodic phase modulation of the BEC wavefunction acted as
a phase grating. Each of the two counterpropagating laser beams
which comprised the standing wave passed through an acousto-optic
modulator (AOM) driven by an arbitrary waveform generator. This
enabled us to control the frequency and phase of each of the beams.
Adding two counterpropagating waves differing in frequency by
$\Delta f$ results essentially in a standing wave that moves with a
velocity $v=\pi\Delta f/G$, where $G$ is the grating vector. The
initial momentum or quasimomentum $\beta$ of the BEC relative to the
standing wave is proportional to $v$. Thus, by varying $\Delta f$,
we could set arbitrarily the value of $\beta$ and also compensate
for the effect of the gravitational acceleration along the standing
wave (the experiments were done in a free-falling frame).\newline

In order to prepare the initial state, the first standing-wave pulse
was relatively long, having a duration of 38 $\mu $s. This pulse
Bragg diffracted the atoms into a superposition of two plane waves
\cite {phillipspaper}: $\left\vert \psi _{0}\right\rangle =\left[
|P=0\rangle +|P=\hbar G\rangle \right] /\sqrt{4\pi }$, where $P$ is
the non-scaled momentum. The second and subsequent pulses of the
standing wave were short enough to be in the Raman-Nath regime.
These pulses diffracted the atoms into a wide spread of momentum
states and enabled the realization of a kicked-rotor system. The
value of the kicking strength $k$ was measured by subjecting the BEC
to one kick and comparing the populations of various diffraction
orders; we used $k\sim 1.4$. By varying the phase of the RF
waveforms driving the AOMs, we were able to shift the position of
the standing wave for the kicked rotor relative to the standing wave
used in the Bragg-state preparation. This is the phase $\gamma $ in
Eqs. (\ref{Nae})-(\ref{ARe}). Finally, in order to probe the
momentum distribution we waited 8 ms and then imaged the atoms in
absorption. Measurements of the initial momentum width of the BEC
using a time-of-flight technique gave an upper bound to $\Delta
\beta $ of 0.1. The slow expansion of the BEC and the finite
resolution of our imaging system made it difficult to measure this
quantity more precisely.\newline

\begin{figure}
  \includegraphics[width=6cm]{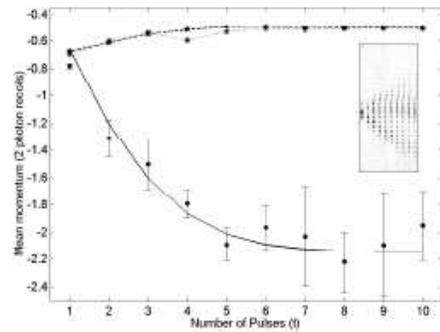}\\
  \caption{Mean momentum vs. kick number $t$ for $k=1.4$,
$\protect\gamma = \protect\pi /2$, and $\protect\beta =0.5$. The
filled circles and error bars are experimental data and the solid
line corresponds to Eq. (\protect\ref{ARe}) ($\Delta \protect\beta
=0.056$). The dashed and dotted lines are the classical mean
momentum $\left\langle p_{t}\right\rangle $ with average taken over
two different initial ensembles in phase space (see text for more
details). The inset shows the time-of-flight images of the BEC vs.
$t$.}\label{fig2}
\end{figure}
We have performed a comprehensive experimental study of the mean
momentum of the BEC as function of several variables. Error bars for
all the data were accurately determined by repeated measurements of
the mean momentum at fixed values of the parameters. The results are
presented in Figs. 1-3 and are compared with the theory above. These
figures show the dependence of the mean momentum on the phase
$\gamma $ for $t=5$ and resonant $\beta =0.5$ (Fig. 1), its
dependence on $t$ for $\gamma =\pi /2$ and $\beta =0.5$ (Fig. 2),
and the dependence of the mean-momentum change on $\beta $ for $t=5$
and $\gamma =\pm \pi /2$ (Fig. 3). The solid line in the figures
corresponds to Eq. (\ref{ANae}) or Eq. (\ref{ARe}). The dashed line
in Figs. 1 and 3 corresponds to the non-averaged theory (\ref{Nae}).
We can see that the experimentally determined mean momentum in all
the figures is well fitted by the theory (\ref{ANae}) or (\ref{ARe})
for the \emph{same} value of the width $\Delta \beta $, $\Delta
\beta =0.056$. This value of $\Delta \beta $ is also consistent with
what can be measured directly using time-of-flight. These facts
indicate good agreement of the experimental results with the
QR-ratchet theory above. Thus, the clear saturation of the mean
momentum in Fig. 2 provides experimental evidence for the
suppression of the resonant ratchet acceleration. The inset in Fig.
2 plots the time-of-flight images of the kicked BEC as time
increases. Note that the distribution of the momentum states is not
symmetric and is weighted towards the negative diffraction orders,
as expected from the mean-momentum values. The finite-width ratchet
effect is similar to the ideal ($\Delta\beta =0$) one in that it
exhibits the current reversal as $\gamma$ is varied (Fig. 1) and it
is pronounced around resonant $\beta =0.5$ (Fig. 3).\newline

\begin{figure}
  \includegraphics[width=6cm]{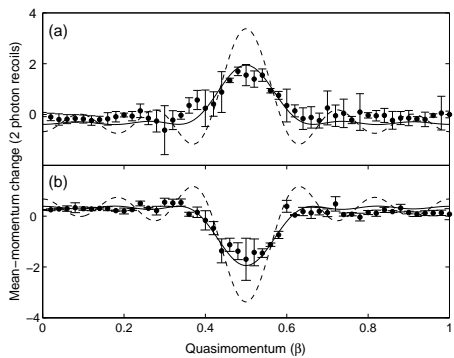}\\
  \caption{Mean-momentum change after $t=5$ kicks vs. $\protect\beta $
for $k=1.4$ and (a) $\protect\gamma =-\protect\pi /2$, (b)
$\protect\gamma = \protect\pi /2$. The filled circles and error bars
are experimental data; the dashed and solid lines correspond,
respectively, to Eqs. (\protect\ref{Nae}) and (\protect\ref{ANae})
($\Delta \protect\beta =0.056$).}\label{fig3}
\end{figure}
It is instructive to compare the behavior of the quantum mean
momentum in Fig. 2 with that of its closest classical analogue. This
is $\langle p_{t}\rangle $, where $p_{t}$ is the classical momentum
at time $t$, given by a standard map, and $\langle \ \rangle $
denotes average over an ensemble of initial conditions
$(p_{0},x_{0})$ whose distribution in phase space is the same as
that featured by the quantum averages in (\ref{ARe}). Thus, the
distribution function of $x_{0}$ is $\phi (x_{0})=\cos
^{2}(x_{0}/2)]/\pi $, equal to the probability density $\left\vert
\psi _{0}(x)\right\vert ^{2}$ for the initial wave packet; $p_{0}$
is distributed like $\langle \hat{p} \rangle _{0}=\langle
\hat{N}\rangle _{0}+\beta ^{\prime }$, where $\langle \hat{N}\rangle
_{0}=-0.5$ and the distribution of $\beta ^{\prime }$ around $ \beta
=0.5$ is the Gaussian $\Gamma _{\beta ,\Delta \beta }(\beta ^{\prime
}) $ above with $\Delta \beta =0.056$. One has $\langle p_{0}\rangle
= \left\langle \langle \hat{p}\rangle _{0}\right\rangle _{r,\Delta
\beta }=0$. The results for $\langle p_{t}\rangle $ are plotted in
Fig. 2 (dashed line), showing a saturation to a value clearly
\emph{different} from the quantum one. In fact, this classical
saturation is \emph{not} due to averaging over $ p_{0}$ or $\beta
^{\prime }$, as in the quantum case, but to the relaxation of the
initial non-uniform distribution $\phi (x_{0})$ to an almost uniform
one in phase space, which is fully chaotic for $k=1.4$. The
transient non-uniformity during the relaxation process causes
$\langle p_{t}\rangle $ to acquire its non-zero saturated value. To
show explicitly that the average over $p_{0}$ is classically not
essential, we plot in Fig. 2 also $\langle p_{t}\rangle $ with the
average taken over an ensemble with constant $p_{0}=0 $ and with
$x_{0}$ distributed as above (dotted line). We see that $\langle
p_{t}\rangle $ saturates to the same value. All this demonstrates
that the experimental results in Fig. 2, as well as those in Figs. 1
and 3, reflect \emph{purely} quantum phenomena for a non-small value
of the effective Planck constant, $\tau /(2\pi )=1$.\newline

In conclusion, we have experimentally realized QR ratchets for
arbitrary quasimomentum $\beta $. These purely quantum ratchets are
unique in their strong dependence on a \emph{conserved quantum}
entity ($\beta $, see Fig. 3) and in their fully \emph{symmetric}
features on which they also depend strongly through the phase
$\gamma $ (Fig. 1). Another remarkable property of the QR ratchets
studied in this work is that they are totally unaffected by the
addition of \emph{arbitrary high harmonics} to the potential $V(x)$,
which usually make $V(x)$ asymmetric. The consideration of general
$\beta $ is necessary to account for the finite width $\Delta \beta
$ of the BEC. We have shown that this width is the main reason for
the suppression of the resonant ratchet acceleration, a distinctive
feature of ideal QR ratchets \cite{qrr,dr}. A fingerprint of this
acceleration for finite $\Delta\beta$ is the pronounced ratchet
effect around resonant $\beta$ (Fig. 3). To increase further the
resonant directed current, $\Delta\beta$ has to be decreased below
the value used in this work. We hope to achieve this in the future
by improving our present experimental setup. Smaller values of
$\Delta \beta $ are also necessary for the realization of more
complex QR ratchets associated, e.g., with free-falling frames
\cite{dr} and/or high-order QRs.\newline

I.D. was partially supported by the Israel Science Foundation (Grant
No. 118/05).


\begin{thebibliography}{99}
\bibitem{qc} \emph{Quantum Chaos, between Order and Disorder}, edited by G.
Casati and B. Chirikov (Cambridge University Press, Cambridge,
1995), and references therein.

\bibitem{qr} F.M. Izrailev, Phys. Rep. \textbf{196}, 299 (1990), and
references therein.

\bibitem{ao} F.L. Moore \textit{et al.}, Phys. Rev. Lett. \textbf{73}, 2974
(1994).

\bibitem{qam} R.M. Godun \textit{et al.}, Phys. Rev. A \textbf{62}, 013411
(2000); S. Schlunk \textit{et al.}, Phys. Rev. Lett. \textbf{90},
054101 (2003); Z.-Y. Ma \textit{et al.}, Phys. Rev. Lett.
\textbf{93}, 164101 (2004).

\bibitem{qam1} G. Behinaein \textit{et al.}, Phys. Rev. Lett. \textbf{97},
244101 (2006).

\bibitem{kp} S. Wimberger, I. Guarneri, and S. Fishman, Nonlinearity \textbf{
16}, 1381 (2003).

\bibitem{kp1} M.B. d'Arcy \textit{et al.}, Phys. Rev. E \textbf{69}, 027201
(2004).

\bibitem{dd} I. Dana and D.L. Dorofeev, Phys. Rev. E \textbf{73}, 026206
(2006); \textit{ibid.} \textbf{74}, 045201(R) (2006).

\bibitem{hqr} C. Ryu \textit{et al.}, Phys. Rev. Lett. \textbf{96}, 160403
(2006).

\bibitem{crd} G.G. Carlo \textit{et al.}, Phys. Rev. A \textbf{74}, 033617
(2006).

\bibitem{qrae} P.H. Jones \textit{et al.}, Phys. Rev. Lett. \textbf{98},
073002 (2007).

\bibitem{qrr} E. Lundh and M. Wallin, Phys. Rev. Lett. \textbf{94}, 110603
(2005).

\bibitem{dr} I. Dana and V. Roitberg, Phys. Rev. E \textbf{76}, 015201(R)
(2007). This paper considers the kicked particle in the presence of
gravity, characterized by coprime integers $(w,\ T)$; the special
case of zero gravity, treated in this Letter, corresponds to $w=0$
and $T=1$.

\bibitem{eqr0} M. Sadgrove \textit{et al.}, Phys. Rev. Lett. \textbf{99},
043002 (2007).

\bibitem{ra} P. Reimann, Phys. Rep. \textbf{361}, 57 (2002); R.D. Astumian
and P. H\"{a}nggi, Phys. Today \textbf{55}, No. 11, 33 (2002).

\bibitem{ra1} C. Robilliard, D. Lucas, and G. Grynberg, Appl. Phys. A
\textbf{75}, 213 (2002); M. Schiavoni \textit{et al.}, Phys. Rev.
Lett. \textbf{90}, 094101 (2003); R. Gommers, M. Brown, and F.
Renzoni, Phys. Rev. A \textbf{75}, 053406 (2007).

\bibitem{cra} H. Schanz, T. Dittrich, and R. Ketzmerick, Phys. Rev. E
\textbf{71}, 026228 (2005).

\bibitem{qra1} J. Gong and P. Brumer, Phys. Rev. Lett. \textbf{97}, 240602
(2006).

\bibitem{Ketterle} K.B. Davis \textit{et al.}, Phys. Rev. Lett. \textbf{75},
3969 (1995).

\bibitem{phillipspaper} M. Kozuma \textit{et al.}, Phys. Rev. Lett. \textbf{
82}, 871 (1999).
\end{thebibliography}
\end{document}